\documentclass[%
reprint,
 showpacs,
 amsmath,amssymb,
 aps,
]{revtex4-1}

\usepackage{graphicx}% Include figure files
\usepackage{dcolumn}% Align table columns on decimal point
\usepackage{bm}% bold math
\usepackage{color}
\begin{document}

\preprint{APS/123-QED}

\title{Cross-over between Magnetic and Electric Edges in Quantum Hall Systems}

\author{Alain Nogaret}
\email{A.R.Nogaret@bath.ac.uk}
\affiliation{Department of Physics, University of Bath, Bath BA2 7AY, UK}

\author{Puja Mondal, Ankip Kumar, Sankalpa Ghosh}
\affiliation{Department of Physics, Indian Institute of Technology Delhi, New Delhi-110016, India}

\author{Harvey Beere, David Ritchie}
\affiliation{Cavendish Laboratory, University of Cambridge, CB3 0HE, UK}

\begin{abstract}
We report on the transition from magnetic edge to electric edge transport in a split magnetic gate device which applies a notch magnetic field to a two-dimensional electron gas.  The gate bias allows tuning the overlap of magnetic and electric edge wavefunctions on the scale of the magnetic length.  Conduction at magnetic edges - in the 2D-bulk - is found to compete with conduction at electric edges until magnetic edges become depleted.  Current lines then move to the electrostatic edges as in the conventional quantum Hall picture.  The conductivity was modelled using the quantum Boltzmann equation in the exact hybrid potential.  The theory predicts the features of the bulk-edge cross-over in good agreement with experiment.
\end{abstract}

%\pacs{}

\maketitle

Spatially varying magnetic fields in two-dimensional electron gases (2DEGs) have revealed novel physics including electrically driven spin resonance ~\cite{Tarucha2008, Nog2007a, Nog2007b}, spin filtering ~\cite{Zhai2002, Xu2001, Liu2006, Anna2009, Papp2001}, current channelling ~\cite{Uzur2004, Lawton2001, Hara2004, Nogaret2000}, magnetic tunnelling barriers and quantum wells ~\cite{Sun1995, Cerchez2007, Wang2009, Song2005, Kubrak1999, Vancura2000, Ramezani2008, Sharma2011} and helical magnetic ordering underpinned by Chern numbers ~\cite{Haldane1995, Young2014}.  Magnetic field gradients support chiral current carrying states ~\cite{Tarasov2010, Schuler2014} which realize the equivalent of conventional edge states in the 2D-bulk of quantum Hall systems.  The recent discovery of exotic electronic states in topological insulators~\cite{Hasan2010} and topological semi-metals~\cite{Xu2016} has revealed the correspondence of non-trivial bulk phases with surface states.  Such discoveries underscore the importance of studying the bulk-edge correspondence in magnetically modulated quantum Hall systems and other topological materials.

%------------------------- Figure1:-------------------------------

\begin{figure}
\includegraphics[width=\linewidth]{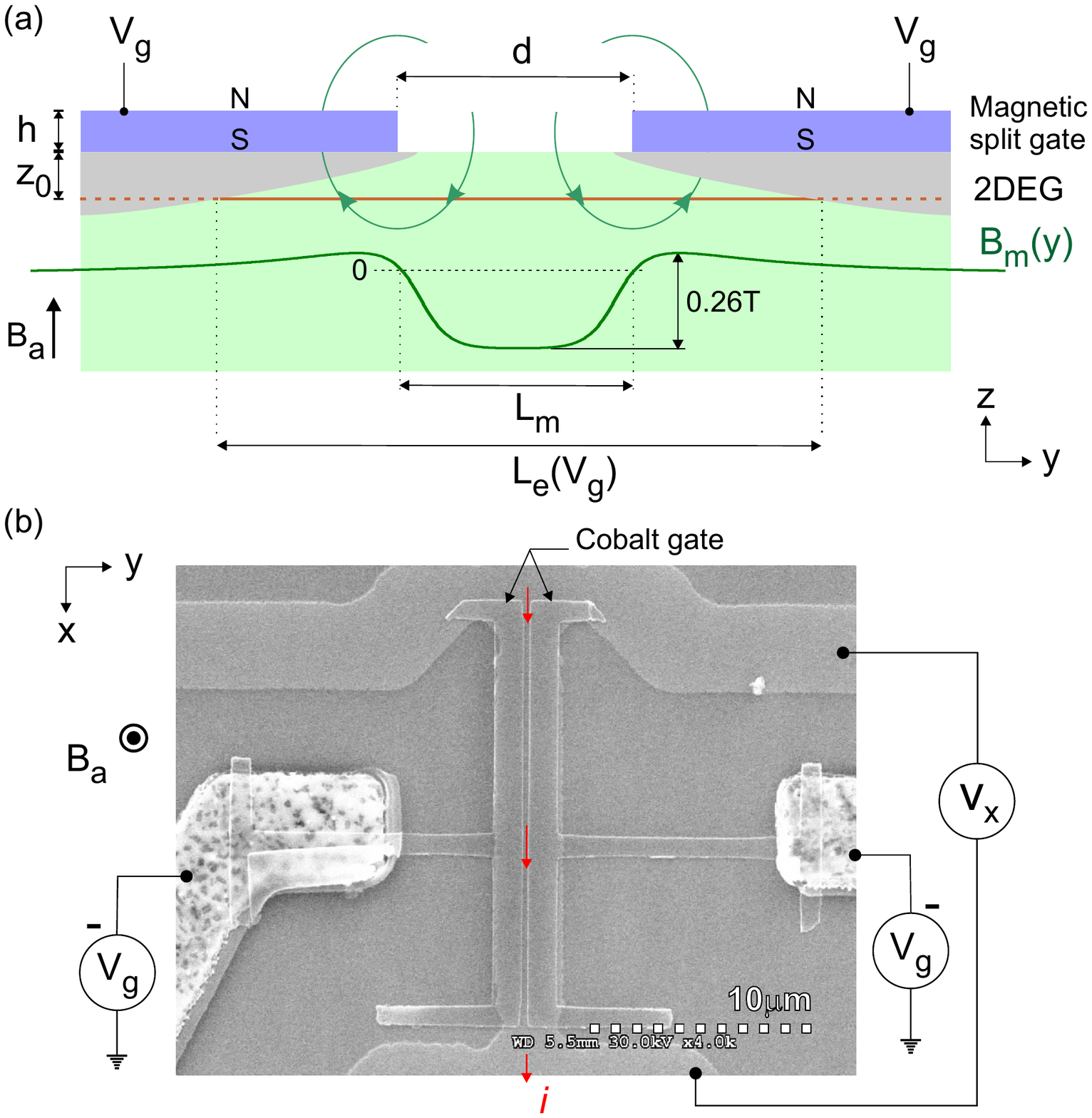}
\caption{\label{fig:fig1}{\it (color online)} (a) The stray magnetic field emanating from the magnetic split gate modulates the 2DEG with perpendicular magnetic field $B_m(y)$.  The central magnetic domain $B_m \le 0$ has constant width, $L_m=274$nm.  A negative gate bias induces equipotential lines (grey area) which deplete the 2DEG and vary its electrical width, $L_e(V_g)$.  (b) SEM image of the device.  $d=200$nm, $h=80$nm, $z_0=30$nm}
\end{figure}

%------------------------------------------------------------------

In this paper, we probe the cross-over of 1D states that carry a current both in the bulk and at the edges of quantum Hall systems.  Bulk 1D states are confined by microscopic magnetic field gradients at the centre of a Hall bar.  By gradually depleting the 2DEG from its etched boundaries towards the centre, we are able to control the overlap of electric and magnetic edge state wavefunctions on the scale of the magnetic length.  We observe the cross-over of magnetic and electric edge states as a reversal in the shift direction of magnetoresistance peaks with decreasing gate bias.  The amplitude of resistance peaks is found to be enhanced near the cross-over as a result of increased interband scattering between magnetic and electric edge subbands.  We model the magnetoresistance using quantum perturbation theory~\cite{Shi2002,Charbonneau1982} in good agreement with observation.  A comparison of magnetoresistance oscillations of magnetic and non-magnetic split gate devices demonstrates that 1D states in the bulk compete with 1D states at the edges to carry the current.  This finding complements the conventional interpretation of the quantum Hall effect.

We synthesized a 30nm deep Al$_{0.33}$Ga$_{0.67}$As/GaAs heterojunction to minimize the decay of the stray magnetic field from micro-magnets fabricated at the surface.  A $\delta$-doped layer, Si: $1.4\times 10^{13}$cm$^{-2}$, was inserted in the top AlGaAs spacer layer, 12.5nm away from the heterojunction, to create excess carriers.  The mobility and electron density were respectively $\mu=3.0\pm0.03\times10^{5}$cm$^{2}$.V$^{-1}$.s$^{-1}$ and $n_s=5.3\pm0.03\times 10^{11}$cm$^{-2}$ at 4K.  The Hall bars were $2\mu$m wide and had voltage probes separated by $18\mu$m.  The Hall bars were gated by a cobalt split gate which applied both electrostatic and magnetic potentials to the 2DEG.  The gates were fabricated using precision electron beam lithography and magnetron sputtering of a 80nm cobalt film. The interval between gates was nominally constant ($d=200$nm) over the length of the Hall channel which resulted in a one-dimensional lateral potential being applied to the 2DEG (Fig.\ref{fig:fig1}).

%------------------------- Figure2:-------------------------------

\begin{figure}
\includegraphics[width=0.95\columnwidth]{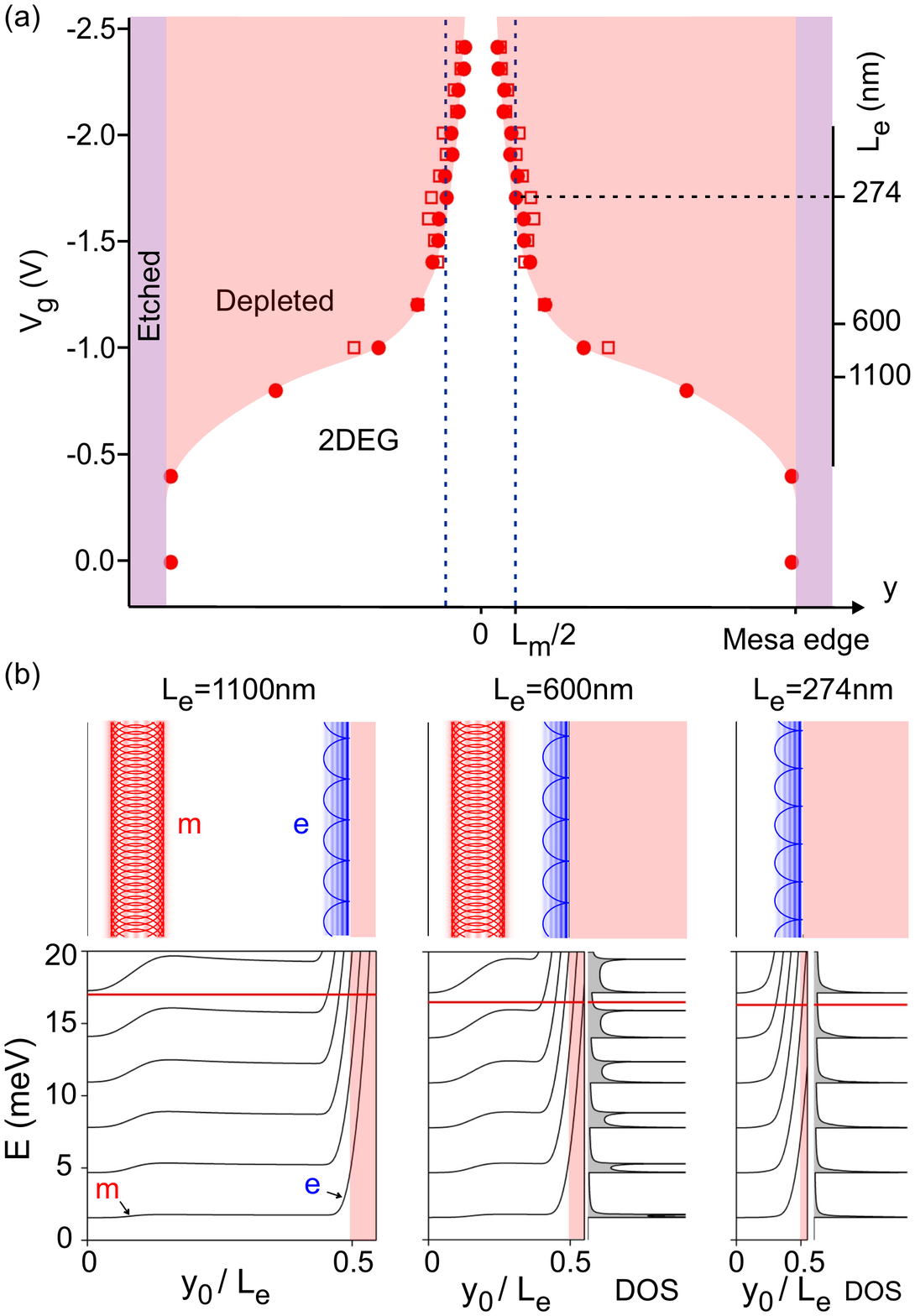}
\caption{\label{fig:fig2}{\it (color online)} (a) Dependence of the electrical width of the channel, $L_e$, on the gate bias estimated from magnetosize effects (hollow squares) and the zero-field resistance (dots).  The depletion region (pink-shaded) crosses over the magnetic edge at $V_g=-1.7$V when $L_e=L_m$ (274nm).  Magnetic edge states are depleted when $V_g \le -1.4$V.  (b) Energy subbands plotted as a function of the position of the centre of oscillator, $y_0$, for $L_e=1100$nm, $600$nm and $274$nm at $B_a$=2T.  Magnetic minibands ($m$) consist of cycloid states (red probability density) centered at $y_0 \approx L_m/2$.  Electric minibands (e) consist of skipping orbits (blue probability density) at $y_0 \approx L_e/2$.  The horizontal red line is the Fermi level.}
\end{figure}

%------------------------------------------------------------------

A quantizing magnetic field $B_a$ was also applied to the 2DEG.  This field magnetized the gate to saturation.  The magnetic poles created on the upper and lower surfaces of the gate induced loops of stray magnetic field which closed through the gap.  As a result, the magnetic modulation perpendicular to the 2DEG had the notch-like profile $B_m(y)$ shown in Fig.\ref{fig:fig1}(a):

\begin{equation}
B_m(y)=-\frac{\mu_0 M_s}{2\pi}\left[f_{0}^{+}(y)-f_{0}^{-}(y)-f_{h}^{+}(y)+f_{h}^{-}(y)\right], \label{eq:eq1}
\end{equation}

\noindent where $f_{z}^{\pm}(y)=\arctan\frac{y \pm d/2}{z_0+z}$~\cite{Nogaret2000}, $z_0=30$nm and $\mu_0 M_s=1.4$T (Co).  The magnetic modulation was negative at the centre of the Hall channel over a width $L_m=274$nm.  This strip was bounded by magnetic field gradients at $y \approx \pm L_m/2$ which confine open orbits in the 2D bulk.  These electron orbits follow lines of constant magnetic field hence run parallel to the electric edge states which follow edge equipotentials.  The 4-terminal resistance was measured at 1.3K.  The probe current was $I=100$nA (13Hz) and the longitudinal voltage $V_x$ was measured using lock-in detection (Fig.\ref{fig:fig1}(b)).

%------------------------- Figure3:-------------------------------

\begin{figure}
\includegraphics[width=\linewidth]{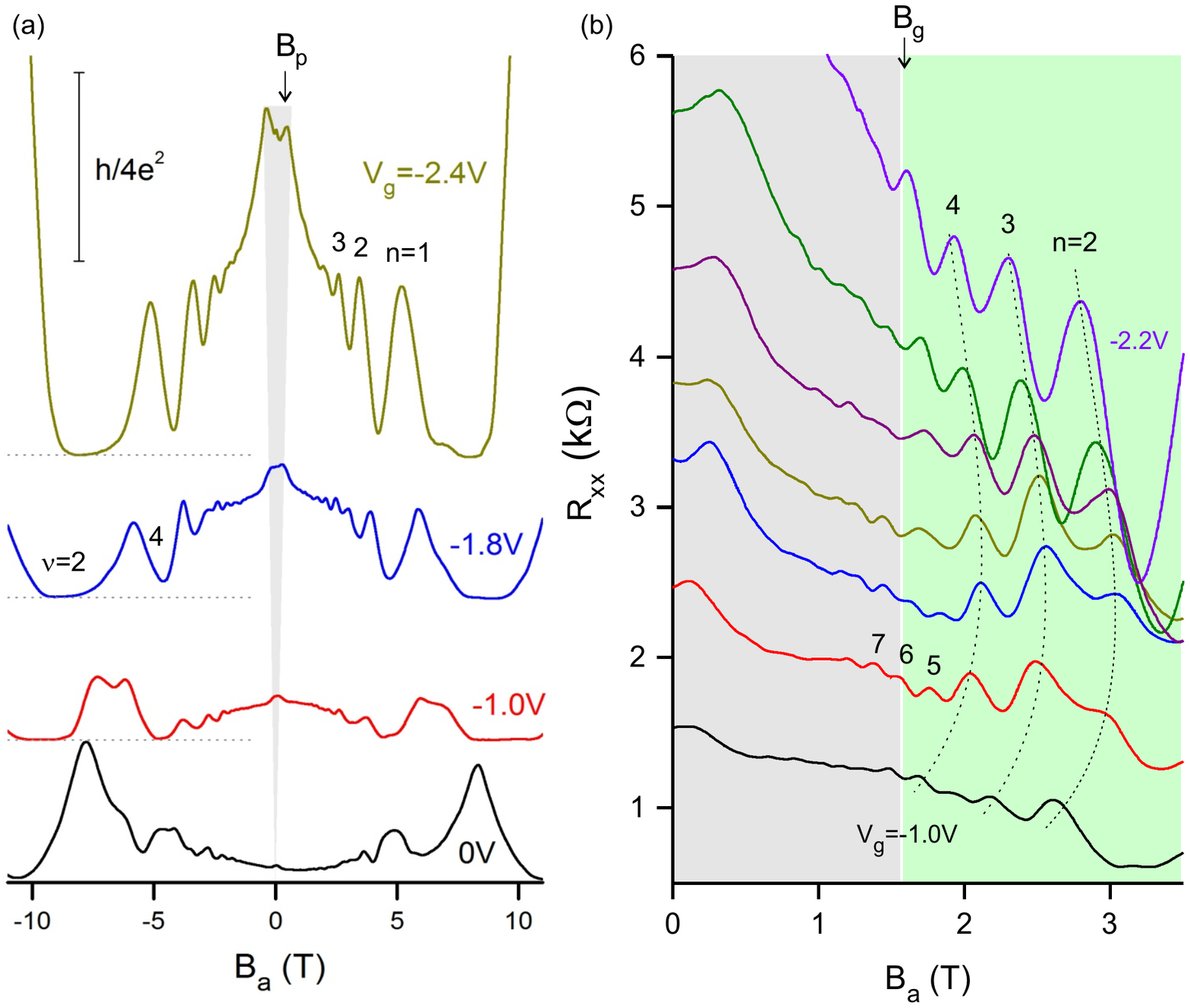}
\caption{\label{fig:fig3}{\it (color online)} (a) Magnetoresistance at fixed gate biases.  $B_p$ is the resistance peak due to boundary scattering.  (b) Magnetoresistance at constant gate bias varying from $V_g=-1.0$V to $-2.2$V in steps of $-0.2$V.  $B_g$ is the critical magnetic field above which energy gaps open between magnetic minibands.  $\nu$: filling fraction, $n$: Landau index.}
\end{figure}

%------------------------------------------------------------------

The gate was biased at a negative potential to gradually deplete the 2DEG.  The channel width decreased from 2$\mu$m to 150nm as the depletion edge moved from the etched boundary ($V_g=0$V) towards the centre of the magnetic notch ($V_g=-2.2$V) through the magnetic edge.  The dependence of the channel width on gate bias, $L_e(V_g)$, is plotted in Fig.\ref{fig:fig2}(a).  This was obtained independently from the gate bias dependence of ballistic magnetosize effects at low magnetic field~\cite{Thornton1989} and the zero field resistance.  Residual backscattering by boundaries gives the resistance peak, $B_p$, in Fig.\ref{fig:fig3}(a).  This peak shifts from 0 to 0.43T as the conducting channel narrows.  The square symbols in Fig.\ref{fig:fig2}(a) indicate the boundary of the electrical channel which we calculated using $L_e=0.55\;\hbar k_F/(eB_p)$ where $k_F=\sqrt{2\pi n_s}$.  $B_p$ was measured for each value of the gate bias (SM~\cite{SM} Table 1).  The dot symbols in Fig.\ref{fig:fig2}(a) show the location of the depletion edge calculated from the 40-fold increase in the zero-field resistance (Fig.\ref{fig:fig3}(a)) as $V_g$ decreases from 0 to -2.4V.  Both methods give highly consistent values for $L_e$ (Fig.\ref{fig:fig2}(a)).  The onset of depletion under the gate occurs at $V_g=-0.4$V.  This figure is in agreement with the capacitive bias $\Delta V_g = e (z_0+\bar{z})\Delta n_s/\epsilon_r\epsilon_0 =-0.35$V required to discharge the 2DEG by $\Delta n_s=-5.3\times10^{11}$cm$^{-2}$.  Here, $\bar{z}=9.5$nm is the Fang-Howard offset of the 2D wavefunction in the potential well of the heterojunction~\cite{Fang1966} and $\epsilon_r$ is the relative dielectric constant of GaAs.  At $V_g = -1.7$V, the depletion edge crosses the magnetic edge at which point $L_e=L_m=274$nm (Fig.\ref{fig:fig2}(a), horizontal dashed line).  Depleting the 2DEG further confines carriers to a narrow strip of nearly uniform magnetic field within the $B_m<0$ magnetic domain.

Fig.\ref{fig:fig2}(b) plots the energy subbands in the hybrid potential as a function of the location of the center of oscillator, $y_0$, relative to the center of the channel.  Magnetic and electric edge subbands are located at $y_0 \approx L_m/2$ and $y_0 \approx L_e/2$ respectively. Gradually depleting the channel increases the overlap of electric and magnetic edge wavefunctions which begins when $L_e$ approaches $L_m$ within a few magnetic lengths ($l_b=\sqrt{\hbar/eB_a}$).  Hybridization typically begins in the last occupied Landau level when $L_e-L_m$=110nm at $B_a=2$T.  The subsequent changes in band structure, DOS, wavefunctions and semiclassical trajectories are shown in Fig.\ref{fig:fig2}(b).  At $L_e=1100$nm ($V_g=-0.9$V) magnetic and electric edges occupy different regions of the 2DEG.  Their wavefunctions begin to overlap at $L_e=600$nm ($V_g=-1.1$V).  The cross-over is complete at $L_e=274$nm ($V_g=-1.7$V).

The gate bias dependence of the magnetoresistance in the region of the cross-over is plotted in Fig.\ref{fig:fig3}.  Magnetoresistance peaks $n=2,3,4$ shift to \textit{higher} magnetic field from $V_g=-1.0$V to -1.4V.  However, this trend reverses at -1.4V.  From $V_g=-1.4$V to -2.2V peaks shift to lower magnetic field.  We interpret this reversal in shift direction as follows.  From $V_g=-1.0$V to -1.4V, the depletion of magnetic minibands causes the DOS maxima to shift lower in energy, from the center to the bottom of magnetic minibands (600nm$\rightarrow274$nm in Fig.\ref{fig:fig2}(b)).  A higher magnetic field is therefore required to keep the top occupied subband aligned with the Fermi level.  The depletion of magnetic minibands explains the anomalous shift of magnetoresistance peaks to higher magnetic field and demonstrates the importance of miniband conduction at the centre of the channel.  At $V_g=-1.4$V magnetic subbands have been depleted leaving only electric edge subbands.  From $V_g=-1.4$V to -2.2V magnetoresistance peaks shift to lower magnetic field due to the monotonic decrease in electron density (SM~\cite{SM} Fig.4).

%------------------------- Figure4:-------------------------------

\begin{figure*}[t]
\includegraphics[width=\linewidth]{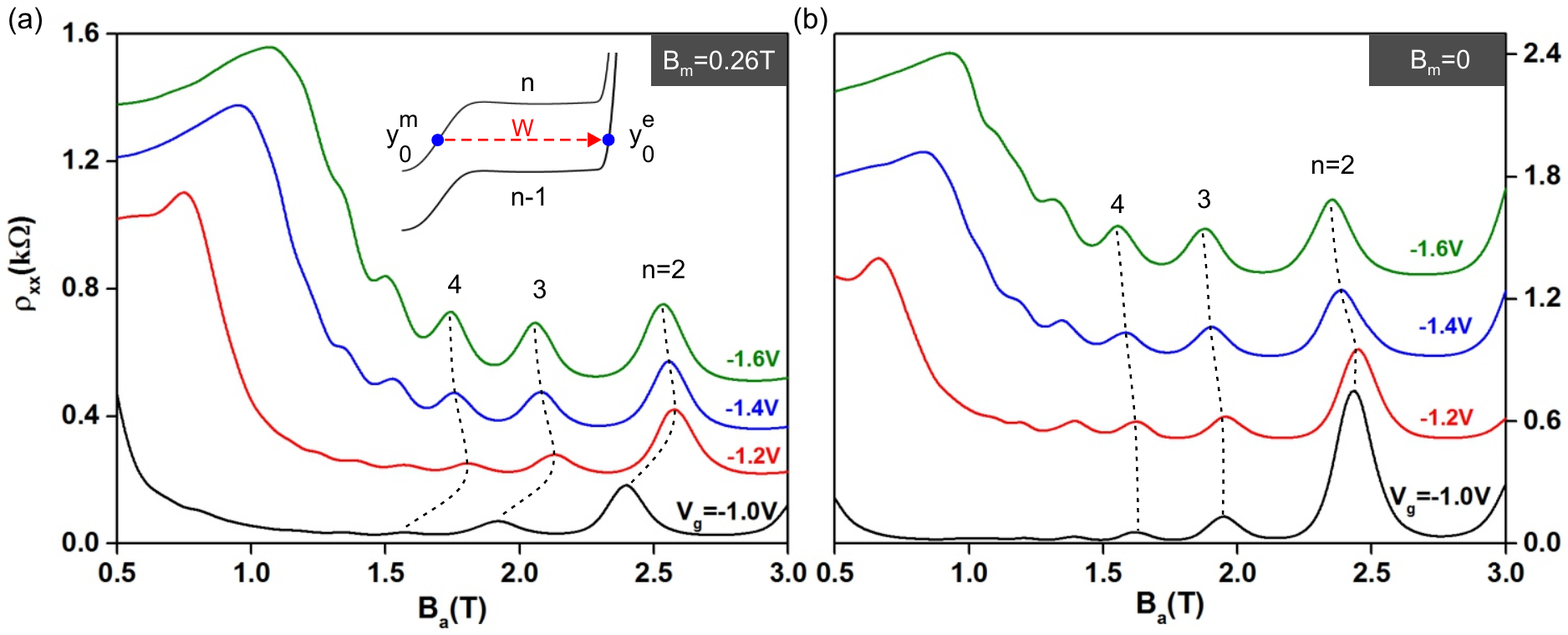}
\caption{\label{fig:fig4}{\it (color online)} (a) Magnetoresistance of the magnetically modulated device calculated near the cross-over at 4K.  The resistance peaks show a reversal in the shift direction with gate bias.  Inset: intersubband elastic transition between centres of oscillator in magnetic subband $n$ and electric subband $n-1$.  (b) Magnetoresistance of the same device gated by a non-magnetic gate.  The resistance peaks shift monotonically.}
\end{figure*}

%------------------------------------------------------------------

The step increase in the amplitude of resistance peaks from $n=7,6,5$ to $n=4,3$ in the $V_g=-1.2$V and $-1.4$V curves of Fig.\ref{fig:fig3}(b) is consistent with the opening of gaps between magnetic minibands.  The $n=7,6,5$ peaks occur in a range of magnetic fields where magnetic minibands overlap at the Fermi level.  Varying $B_a$ changes the miniband overlap and gives the weak amplitude oscillations $n=7,6,5$~\cite{Shi2002,Edmonds2001}.  Once $B_a>(n+1/2)B_m$, energy gaps open between magnetic minibands at the Fermi level (SM~\cite{SM} Fig.5).  By writing that magnetic miniband $n$ crosses the Fermi level as $2n+1=(k_F l_b)^2$, once obtains the critical magnetic field at which gaps open as $B_{g}=\sqrt{B_m B_F}$ where $B_F=\hbar k_F^2/2e$.  Using $B_m=0.26T$ for the amplitude of the magnetic notch, one finds $B_{g}=1.5$T.  This result is in excellent agreement with the onset of the larger amplitude oscillations observed in Fig.\ref{fig:fig3}(b).

We then modelled the bulk-edge cross-over by calculating the conductivity tensor within the linear response theory~\cite{Shi2002,Charbonneau1982}.  The energies $E_n$ and wavefunctions $\varphi_n$ are solutions of the one-dimensional Schr\"{o}dinger equation:

\small
\begin{equation}
\left[-\frac{\partial^2 }{\partial \tilde{y}^2}+\left( \tilde{y}_0+A(\tilde{y})/B_a\right)^2+\tilde{V}(y)\right]\varphi_n(\tilde{y})=\tilde{E}_{n}(y_0)\varphi_n(\tilde{y}), \label{eq:eq2}
\end{equation}
\normalsize

\noindent where $y_{0}=k_x l_b^2$ is the position of the centre of oscillator, $\tilde{y}_{0}=y_0/l_b$, $\tilde{y}=y/l_b$, $\tilde{E}_n(y_0)=2m^*E_n(y_0)l_b^2/\hbar^2$, $\tilde{V}(y)=2m^*V(y)l_b^2/\hbar^2$ and $m^*$ is the electron effective mass.  The lateral potential comprises the vector potential of both homogeneous and inhomogeneous magnetic fields:

\begin{equation}
A(y)=-y B_a - \int_{-\infty}^{y} dy' B_m(y'),
\label{eq:eq3}
\end{equation}

\noindent and the scalar potential of the split gate, which we model as a quantum well of tuneable width, $L_e$:

\begin{equation}
V(y)= \left\{ \begin{array}{ll}
V_0 & \mbox{\;\;\;\; $|y| \ge L_e(V_g)/2$}  \\
0 & \mbox{\;\;\;\; otherwise}
\end{array}
\right. ,
\label{eq:eq4}
\end{equation}

\noindent bound by potential barriers of finite height, $V_0=700$meV ~\cite{Reed1988}.  We obtained the energies and eigenfunctions by solving Eq.\ref{eq:eq2} numerically using a two point boundary relaxation method~\cite{Press1994}.  The hybrid potential was switched on adiabatically and the solutions of Eq.\ref{eq:eq2} were allowed to relax within the constraints set by the symmetry and boundary conditions on the wavefunction.  The resulting energy dispersion curves $E_n(y_0)$ are plotted in Fig.\ref{fig:fig2}(b).  Disorder introduces Gaussian broadening of energy levels, $\delta(E-E_\eta) \rightarrow P(E-E_\eta)$ whose linewidth $\Gamma=0.6$meV we fitted from the magnetoresistance peaks in Fig.\ref{fig:fig3}.

The density of states is obtained by summing over quantum states $\eta \equiv \{n,k_x\}$:

\small
\begin{equation}
D(E)= \frac{m*}{\pi\hbar^2} \hbar \omega_c \sum_{\eta} P(E-E_\eta). \label{eq:eq5a}
\end{equation}
\normalsize

\noindent The magneto-oscillations of the DOS exhibit a change in shift direction of DOS peaks as $V_g$ decreases (SM~\cite{SM} Fig.6).  Theory predicts the shift reversal to occur at $V_g=-1.4$V which agrees well with experiment.  The DOS oscillations at $V_g=-1.0$V display a series of satellite peaks which are absent in the $-1.4$V and $-1.6$V curves.  These are the van Hove singularities of magnetic minibands.

The quantum conductivity tensor~\cite{Shi2002,Charbonneau1982}, $\sigma_{\alpha\beta}=\sigma_{\alpha\beta}^{d}+\sigma_{\alpha\beta}^{nd}+\sigma_{\alpha\beta}^{c}$ ($\alpha,\beta \equiv x,y$) was obtained by calculating contributions of the band conductivity, $\sigma_{\alpha\beta}^{d}$, which depends on the diagonal matrix element of the electron drift velocity $\langle \eta|v_\alpha |\eta \rangle $; the non-diagonal band conductivity, $\sigma_{\alpha\beta}^{nd}$, which depends on $\langle \eta|v_\alpha |\eta' \rangle $; and the collisional conductivity, $\sigma_{\alpha\beta}^{c}$.  Band conduction is perpendicular to the edges therefore $\langle v_y \rangle_\eta=0$.  It follows that all $\sigma_{\alpha\beta}^{d}$ vanish except $\sigma_{xx}^{d}$.  The symmetry of the gate potential about $y=0$ implies $\sigma_{xx}^{nd}=0$ and $\sigma_{yy}^{nd}$, $\sigma_{xy}^{nd}$, $\sigma_{yx}^{nd}$ have finite values.  Disorder scattering is isotropic, hence $\sigma_{xx}^{c}=\sigma_{yy}^{c}$ and $\sigma_{xy}^{c}=\sigma_{yx}^{c}=0$.  The elastic scattering rate assisted by background impurities is:

\small
\begin{equation}
W_{\eta,\eta'}=\frac{2 \pi}{\hbar} N_I \sum_{\bf{q}} \frac{U_0^2}{q_x^2+q_y^2+k_s^2}\left| \langle \eta|e^{i\bf{qr}}|\eta' \rangle \right|^2 \delta \left(E_\eta - E_{\eta'}\right), \label{eq:eq7}
\end{equation}
\normalsize

\noindent where $N_I=1.4\times10^{13}$cm$^{-2}$ is the nominal dopant density, $U_0=e^2/2\epsilon_r \epsilon_0$ the Coulomb potential, and $k_s=2m^*e^2/(\epsilon_r\epsilon_0\hbar^2)$ is the inverse screening length.

The peculiar band structure of the hybrid potential allows elastic scattering to occur between magnetic and electric subbands (inset to Fig.\ref{fig:fig4}).  We find the scattering rate $W_{(n-1,y_0^m) \rightarrow (n,y_0^e)} \propto e^{-\gamma}$ increases exponentially as the overlap of edge wavefunctions increases.  $\gamma = |y_0^m-y_0^e|^2/l_b^2$ where $y_0^m$ and $y_0^e$ are the magnetic and electric centres of oscillator.  A dependence of the collisional conductivity on $V_g$ is therefore expected as the interband scattering rate will increase near the cross-over.  We calculated the collisional conductivity as~\cite{Shi2002}:

% collisional conductivity
\small
\begin{equation}
\sigma_{xx}^{c} = \frac{\beta e^2}{A}\sum_{\eta,\eta'}\iint dE dE' D_{\eta,\eta'}(E,E')W_{\eta,\eta'}(\langle x \rangle_\eta-\langle x \rangle_{\eta'})^2, \label{eq:eq8}
\end{equation}
\normalsize

\noindent where $D_{\eta,\eta'}(E,E') = P(E-E_\eta)P(E'-E_{\eta'})f(E)[1-f(E')]$ and $f(E)$ is the Fermi function. The resistivity is then computed as:
$\rho_{xx}=\sigma_{yy}/(\sigma_{xx}\sigma_{yy}-\sigma_{xy}\sigma_{yx})$.

The theoretical magnetoresistance of the magnetic split gate device is plotted in Fig.\ref{fig:fig4}(a).  The magnetoresistance of the equivalent device with non-magnetic gates is shown in Fig.\ref{fig:fig4}(b).  A comparison of both plots demonstrates that the reversal in shift direction of magnetoresistance peaks is the signature feature of the magnetic modulation.  The reversal is due to the depletion of magnetic minibands and indicates the importance of miniband conduction in the 2D-bulk.  One also notes that the resistance peaks of the $V_g=-1.0$V curve occur at a lower magnetic field when $B_m$ is finite (Fig.\ref{fig:fig4}(a)).  Clearly conduction occurs through states of lower energy than conventional Landau levels which are the magnetic subbands at the centre of the channel.  If bulk conduction were negligible, as prescribed by the model of the quantum Hall effect, the magnetoresistance of electrically and electrically+magnetically modulated channels would be identical.  This is not the case.  The approach of the edge cross-over is further characterized by a small increase in amplitude of resistance peaks when $V_g$ decreases from -1.0V to -1.4V. The theory shows this increase is due to a four-fold enhancement of the collisional conductivity by interband scattering (inset Fig.\ref{fig:fig4}(a)).  This qualitatively agrees with observation (Fig.\ref{fig:fig3}(b)).  The theory also predicts a low resistance peak at 0.7T in the $V_g=-1.2$V curve.  Like $B_p$ this peak shifts to higher magnetic field as the lateral electrostatic confinement increases.  Semi-classical arguments based on Boltzmann's equation~\cite{Pippard1989} require the edge boundaries to be diffuse for magnetosize effects to produce any effect in the resistance.  Here our quantum mechanical calculation shows that a robust resistance peak may still occur even when edge boundaries are specular.  This peak arises from the transition from diffusion to collision dominated transport and occurs at a magnetic field 3.5 times larger than $B_p$.  Screening effects~\cite{Chklovskii1992} would shift this peak to lower magnetic field but would not generally modify the amplitude or position of magnetoresistance peaks~\cite{Siddiki2004}.

In summary, magnetically modulated quantum Hall systems allow current carrying states to form in the 2D-bulk.  These states compete with electric edges states at the boundaries of the Hall channel giving an anomalous shift in magnetoresistance peaks as a function of gate bias.  The study of low dimensional electron states in the bulk complements the model of the quantum Hall effect and shows that oscillations of the density of states are a powerful tool for probing bulk electronic phases in quantum Hall systems and topological semimetals~\cite{Potter2014}.

\begin{acknowledgments}
This work was supported by a UGC-UKIERI 2013-14/081 thematic award.
\end{acknowledgments}

\bibliography{Magwire}

\begin{thebibliography}{38}%
\makeatletter
\providecommand \@ifxundefined [1]{%
 \@ifx{#1\undefined}
}%
\providecommand \@ifnum [1]{%
 \ifnum #1\expandafter \@firstoftwo
 \else \expandafter \@secondoftwo
 \fi
}%
\providecommand \@ifx [1]{%
 \ifx #1\expandafter \@firstoftwo
 \else \expandafter \@secondoftwo
 \fi
}%
\providecommand \natexlab [1]{#1}%
\providecommand \enquote  [1]{``#1''}%
\providecommand \bibnamefont  [1]{#1}%
\providecommand \bibfnamefont [1]{#1}%
\providecommand \citenamefont [1]{#1}%
\providecommand \href@noop [0]{\@secondoftwo}%
\providecommand \href [0]{\begingroup \@sanitize@url \@href}%
\providecommand \@href[1]{\@@startlink{#1}\@@href}%
\providecommand \@@href[1]{\endgroup#1\@@endlink}%
\providecommand \@sanitize@url [0]{\catcode `\\12\catcode `\$12\catcode
  `\&12\catcode `\#12\catcode `\^12\catcode `\_12\catcode `\%12\relax}%
\providecommand \@@startlink[1]{}%
\providecommand \@@endlink[0]{}%
\providecommand \url  [0]{\begingroup\@sanitize@url \@url }%
\providecommand \@url [1]{\endgroup\@href {#1}{\urlprefix }}%
\providecommand \urlprefix  [0]{URL }%
\providecommand \Eprint [0]{\href }%
\providecommand \doibase [0]{http://dx.doi.org/}%
\providecommand \selectlanguage [0]{\@gobble}%
\providecommand \bibinfo  [0]{\@secondoftwo}%
\providecommand \bibfield  [0]{\@secondoftwo}%
\providecommand \translation [1]{[#1]}%
\providecommand \BibitemOpen [0]{}%
\providecommand \bibitemStop [0]{}%
\providecommand \bibitemNoStop [0]{.\EOS\space}%
\providecommand \EOS [0]{\spacefactor3000\relax}%
\providecommand \BibitemShut  [1]{\csname bibitem#1\endcsname}%
\let\auto@bib@innerbib\@empty
%</preamble>
\bibitem [{\citenamefont {Pioro-Ladri\`{e}re}\ \emph
  {et~al.}(2008)\citenamefont {Pioro-Ladri\`{e}re}, \citenamefont {Obata},
  \citenamefont {Tokura}, \citenamefont {Shin}, \citenamefont {Kubo},
  \citenamefont {Yoshida}, \citenamefont {Taniyama},\ and\ \citenamefont
  {Tarucha}}]{Tarucha2008}%
  \BibitemOpen
  \bibfield  {author} {\bibinfo {author} {\bibfnamefont {M.}~\bibnamefont
  {Pioro-Ladri\`{e}re}}, \bibinfo {author} {\bibfnamefont {T.}~\bibnamefont
  {Obata}}, \bibinfo {author} {\bibfnamefont {Y.}~\bibnamefont {Tokura}},
  \bibinfo {author} {\bibfnamefont {Y.~S.}\ \bibnamefont {Shin}}, \bibinfo
  {author} {\bibfnamefont {T.}~\bibnamefont {Kubo}}, \bibinfo {author}
  {\bibfnamefont {K.}~\bibnamefont {Yoshida}}, \bibinfo {author} {\bibfnamefont
  {T.}~\bibnamefont {Taniyama}}, \ and\ \bibinfo {author} {\bibfnamefont
  {S.}~\bibnamefont {Tarucha}},\ }\href@noop {} {\bibfield  {journal} {\bibinfo
   {journal} {Nature Physics}\ }\textbf {\bibinfo {volume} {4}},\ \bibinfo
  {pages} {776} (\bibinfo {year} {2008})}\BibitemShut {NoStop}%
\bibitem [{\citenamefont {Nogaret}\ and\ \citenamefont
  {Peeters}(2007)}]{Nog2007a}%
  \BibitemOpen
  \bibfield  {author} {\bibinfo {author} {\bibfnamefont {A.}~\bibnamefont
  {Nogaret}}\ and\ \bibinfo {author} {\bibfnamefont {F.~M.}\ \bibnamefont
  {Peeters}},\ }\href@noop {} {\bibfield  {journal} {\bibinfo  {journal}
  {Phys.Rev.B}\ }\textbf {\bibinfo {volume} {76}},\ \bibinfo {pages} {075311}
  (\bibinfo {year} {2007})}\BibitemShut {NoStop}%
\bibitem [{\citenamefont {Nogaret}\ \emph {et~al.}(2007)\citenamefont
  {Nogaret}, \citenamefont {Lambert},\ and\ \citenamefont
  {Peeters}}]{Nog2007b}%
  \BibitemOpen
  \bibfield  {author} {\bibinfo {author} {\bibfnamefont {A.}~\bibnamefont
  {Nogaret}}, \bibinfo {author} {\bibfnamefont {N.~J.}\ \bibnamefont
  {Lambert}}, \ and\ \bibinfo {author} {\bibfnamefont {F.~M.}\ \bibnamefont
  {Peeters}},\ }\href@noop {} {\bibfield  {journal} {\bibinfo  {journal}
  {Phys.Rev.B}\ }\textbf {\bibinfo {volume} {76}},\ \bibinfo {pages} {075312}
  (\bibinfo {year} {2007})}\BibitemShut {NoStop}%
\bibitem [{\citenamefont {Zhai}\ \emph {et~al.}(2002)\citenamefont {Zhai},
  \citenamefont {Guo},\ and\ \citenamefont {Gu}}]{Zhai2002}%
  \BibitemOpen
  \bibfield  {author} {\bibinfo {author} {\bibfnamefont {F.}~\bibnamefont
  {Zhai}}, \bibinfo {author} {\bibfnamefont {Y.}~\bibnamefont {Guo}}, \ and\
  \bibinfo {author} {\bibfnamefont {B.~L.}\ \bibnamefont {Gu}},\ }\href@noop {}
  {\bibfield  {journal} {\bibinfo  {journal} {Phys.Rev.B}\ }\textbf {\bibinfo
  {volume} {66}},\ \bibinfo {pages} {125305} (\bibinfo {year}
  {2002})}\BibitemShut {NoStop}%
\bibitem [{\citenamefont {Xu}\ and\ \citenamefont {Okada}(2001)}]{Xu2001}%
  \BibitemOpen
  \bibfield  {author} {\bibinfo {author} {\bibfnamefont {H.~Z.}\ \bibnamefont
  {Xu}}\ and\ \bibinfo {author} {\bibfnamefont {Y.}~\bibnamefont {Okada}},\
  }\href@noop {} {\bibfield  {journal} {\bibinfo  {journal} {Appl.Phys.Lett.}\
  }\textbf {\bibinfo {volume} {79}},\ \bibinfo {pages} {3119} (\bibinfo {year}
  {2001})}\BibitemShut {NoStop}%
\bibitem [{\citenamefont {Liu}\ \emph {et~al.}(2006)\citenamefont {Liu},
  \citenamefont {Deng}, \citenamefont {Xia}, \citenamefont {Zhang},\ and\
  \citenamefont {Ma}}]{Liu2006}%
  \BibitemOpen
  \bibfield  {author} {\bibinfo {author} {\bibfnamefont {J.-F.}\ \bibnamefont
  {Liu}}, \bibinfo {author} {\bibfnamefont {W.-J.}\ \bibnamefont {Deng}},
  \bibinfo {author} {\bibfnamefont {K.}~\bibnamefont {Xia}}, \bibinfo {author}
  {\bibfnamefont {C.}~\bibnamefont {Zhang}}, \ and\ \bibinfo {author}
  {\bibfnamefont {Z.}~\bibnamefont {Ma}},\ }\href@noop {} {\bibfield  {journal}
  {\bibinfo  {journal} {Phys.Rev.B}\ }\textbf {\bibinfo {volume} {73}},\
  \bibinfo {pages} {155309} (\bibinfo {year} {2006})}\BibitemShut {NoStop}%
\bibitem [{\citenamefont {Dell~Anna}\ and\ \citenamefont
  {De~Martino}(2009)}]{Anna2009}%
  \BibitemOpen
  \bibfield  {author} {\bibinfo {author} {\bibfnamefont {L.}~\bibnamefont
  {Dell~Anna}}\ and\ \bibinfo {author} {\bibfnamefont {A.}~\bibnamefont
  {De~Martino}},\ }\href@noop {} {\bibfield  {journal} {\bibinfo  {journal}
  {Phys.Rev.B}\ }\textbf {\bibinfo {volume} {80}},\ \bibinfo {pages} {155416}
  (\bibinfo {year} {2009})}\BibitemShut {NoStop}%
\bibitem [{\citenamefont {Papp}\ and\ \citenamefont
  {Peeters}(2001)}]{Papp2001}%
  \BibitemOpen
  \bibfield  {author} {\bibinfo {author} {\bibfnamefont {G.}~\bibnamefont
  {Papp}}\ and\ \bibinfo {author} {\bibfnamefont {F.~M.}\ \bibnamefont
  {Peeters}},\ }\href@noop {} {\bibfield  {journal} {\bibinfo  {journal}
  {Appl.Phys.Lett.}\ }\textbf {\bibinfo {volume} {78}},\ \bibinfo {pages}
  {2184} (\bibinfo {year} {2001})}\BibitemShut {NoStop}%
\bibitem [{\citenamefont {Uzur}\ \emph {et~al.}(2004)\citenamefont {Uzur},
  \citenamefont {Nogaret}, \citenamefont {Beere}, \citenamefont {Ritchie},
  \citenamefont {Marrows},\ and\ \citenamefont {Hickey}}]{Uzur2004}%
  \BibitemOpen
  \bibfield  {author} {\bibinfo {author} {\bibfnamefont {D.}~\bibnamefont
  {Uzur}}, \bibinfo {author} {\bibfnamefont {A.}~\bibnamefont {Nogaret}},
  \bibinfo {author} {\bibfnamefont {H.~E.}\ \bibnamefont {Beere}}, \bibinfo
  {author} {\bibfnamefont {D.~A.}\ \bibnamefont {Ritchie}}, \bibinfo {author}
  {\bibfnamefont {C.~H.}\ \bibnamefont {Marrows}}, \ and\ \bibinfo {author}
  {\bibfnamefont {B.~J.}\ \bibnamefont {Hickey}},\ }\href@noop {} {\bibfield
  {journal} {\bibinfo  {journal} {Phys.Rev.B}\ }\textbf {\bibinfo {volume}
  {69}},\ \bibinfo {pages} {241301} (\bibinfo {year} {2004})}\BibitemShut
  {NoStop}%
\bibitem [{\citenamefont {Lawton}\ \emph {et~al.}(2001)\citenamefont {Lawton},
  \citenamefont {Nogaret}, \citenamefont {Bending}, \citenamefont {Maude},
  \citenamefont {Portal},\ and\ \citenamefont {Henini}}]{Lawton2001}%
  \BibitemOpen
  \bibfield  {author} {\bibinfo {author} {\bibfnamefont {D.~N.}\ \bibnamefont
  {Lawton}}, \bibinfo {author} {\bibfnamefont {A.}~\bibnamefont {Nogaret}},
  \bibinfo {author} {\bibfnamefont {S.~J.}\ \bibnamefont {Bending}}, \bibinfo
  {author} {\bibfnamefont {D.~K.}\ \bibnamefont {Maude}}, \bibinfo {author}
  {\bibfnamefont {J.~C.}\ \bibnamefont {Portal}}, \ and\ \bibinfo {author}
  {\bibfnamefont {M.}~\bibnamefont {Henini}},\ }\href@noop {} {\bibfield
  {journal} {\bibinfo  {journal} {Phys.Rev.B}\ }\textbf {\bibinfo {volume}
  {64}},\ \bibinfo {pages} {033312} (\bibinfo {year} {2001})}\BibitemShut
  {NoStop}%
\bibitem [{\citenamefont {Hara}\ \emph {et~al.}(2004)\citenamefont {Hara},
  \citenamefont {Endo}, \citenamefont {Katsumoto},\ and\ \citenamefont
  {Iye}}]{Hara2004}%
  \BibitemOpen
  \bibfield  {author} {\bibinfo {author} {\bibfnamefont {M.}~\bibnamefont
  {Hara}}, \bibinfo {author} {\bibfnamefont {A.}~\bibnamefont {Endo}}, \bibinfo
  {author} {\bibfnamefont {S.}~\bibnamefont {Katsumoto}}, \ and\ \bibinfo
  {author} {\bibfnamefont {Y.}~\bibnamefont {Iye}},\ }\href@noop {} {\bibfield
  {journal} {\bibinfo  {journal} {Phys.Rev.B}\ }\textbf {\bibinfo {volume}
  {69}},\ \bibinfo {pages} {153304} (\bibinfo {year} {2004})}\BibitemShut
  {NoStop}%
\bibitem [{\citenamefont {Nogaret}\ \emph {et~al.}(2000)\citenamefont
  {Nogaret}, \citenamefont {Bending},\ and\ \citenamefont
  {Henini}}]{Nogaret2000}%
  \BibitemOpen
  \bibfield  {author} {\bibinfo {author} {\bibfnamefont {A.}~\bibnamefont
  {Nogaret}}, \bibinfo {author} {\bibfnamefont {S.~J.}\ \bibnamefont
  {Bending}}, \ and\ \bibinfo {author} {\bibfnamefont {M.}~\bibnamefont
  {Henini}},\ }\href@noop {} {\bibfield  {journal} {\bibinfo  {journal}
  {Phys.Rev.Lett.}\ }\textbf {\bibinfo {volume} {84}},\ \bibinfo {pages} {2231}
  (\bibinfo {year} {2000})}\BibitemShut {NoStop}%
\bibitem [{\citenamefont {Sun}\ \emph {et~al.}(1995)\citenamefont {Sun},
  \citenamefont {Kirczenow}, \citenamefont {Sachrajda},\ and\ \citenamefont
  {Feng}}]{Sun1995}%
  \BibitemOpen
  \bibfield  {author} {\bibinfo {author} {\bibfnamefont {Y.}~\bibnamefont
  {Sun}}, \bibinfo {author} {\bibfnamefont {G.}~\bibnamefont {Kirczenow}},
  \bibinfo {author} {\bibfnamefont {A.~S.}\ \bibnamefont {Sachrajda}}, \ and\
  \bibinfo {author} {\bibfnamefont {Y.}~\bibnamefont {Feng}},\ }\href@noop {}
  {\bibfield  {journal} {\bibinfo  {journal} {J.Appl.Phys.}\ }\textbf {\bibinfo
  {volume} {77}},\ \bibinfo {pages} {6361} (\bibinfo {year}
  {1995})}\BibitemShut {NoStop}%
\bibitem [{\citenamefont {Cerchez}\ \emph {et~al.}(2007)\citenamefont
  {Cerchez}, \citenamefont {Hugger}, \citenamefont {Heinzel},\ and\
  \citenamefont {Schulz}}]{Cerchez2007}%
  \BibitemOpen
  \bibfield  {author} {\bibinfo {author} {\bibfnamefont {M.}~\bibnamefont
  {Cerchez}}, \bibinfo {author} {\bibfnamefont {S.}~\bibnamefont {Hugger}},
  \bibinfo {author} {\bibfnamefont {T.}~\bibnamefont {Heinzel}}, \ and\
  \bibinfo {author} {\bibfnamefont {N.}~\bibnamefont {Schulz}},\ }\href@noop {}
  {\bibfield  {journal} {\bibinfo  {journal} {Phys.Rev.B}\ }\textbf {\bibinfo
  {volume} {75}},\ \bibinfo {pages} {035341} (\bibinfo {year}
  {2007})}\BibitemShut {NoStop}%
\bibitem [{\citenamefont {Wang}\ \emph {et~al.}(2009)\citenamefont {Wang},
  \citenamefont {Chen}, \citenamefont {Zhang}, \citenamefont {Chen},
  \citenamefont {Yang}, \citenamefont {Yin},\ and\ \citenamefont
  {Bai}}]{Wang2009}%
  \BibitemOpen
  \bibfield  {author} {\bibinfo {author} {\bibfnamefont {Y.}~\bibnamefont
  {Wang}}, \bibinfo {author} {\bibfnamefont {N.~F.}\ \bibnamefont {Chen}},
  \bibinfo {author} {\bibfnamefont {X.~W.}\ \bibnamefont {Zhang}}, \bibinfo
  {author} {\bibfnamefont {X.~F.}\ \bibnamefont {Chen}}, \bibinfo {author}
  {\bibfnamefont {X.~L.}\ \bibnamefont {Yang}}, \bibinfo {author}
  {\bibfnamefont {Z.~G.}\ \bibnamefont {Yin}}, \ and\ \bibinfo {author}
  {\bibfnamefont {Y.~M.}\ \bibnamefont {Bai}},\ }\href@noop {} {\bibfield
  {journal} {\bibinfo  {journal} {Phys.Lett.A}\ }\textbf {\bibinfo {volume}
  {373}},\ \bibinfo {pages} {1983} (\bibinfo {year} {2009})}\BibitemShut
  {NoStop}%
\bibitem [{\citenamefont {Song}\ \emph {et~al.}(2005)\citenamefont {Song},
  \citenamefont {Bird},\ and\ \citenamefont {Ochiai}}]{Song2005}%
  \BibitemOpen
  \bibfield  {author} {\bibinfo {author} {\bibfnamefont {J.~F.}\ \bibnamefont
  {Song}}, \bibinfo {author} {\bibfnamefont {J.~P.}\ \bibnamefont {Bird}}, \
  and\ \bibinfo {author} {\bibfnamefont {Y.}~\bibnamefont {Ochiai}},\
  }\href@noop {} {\bibfield  {journal} {\bibinfo  {journal} {Appl.Phys.Lett.}\
  }\textbf {\bibinfo {volume} {86}},\ \bibinfo {pages} {062106} (\bibinfo
  {year} {2005})}\BibitemShut {NoStop}%
\bibitem [{\citenamefont {Kubrak}\ \emph {et~al.}(1999)\citenamefont {Kubrak},
  \citenamefont {Rahman}, \citenamefont {Gallagher}, \citenamefont {Main},
  \citenamefont {Henini}, \citenamefont {Marrows},\ and\ \citenamefont
  {Howson}}]{Kubrak1999}%
  \BibitemOpen
  \bibfield  {author} {\bibinfo {author} {\bibfnamefont {V.}~\bibnamefont
  {Kubrak}}, \bibinfo {author} {\bibfnamefont {F.}~\bibnamefont {Rahman}},
  \bibinfo {author} {\bibfnamefont {B.~L.}\ \bibnamefont {Gallagher}}, \bibinfo
  {author} {\bibfnamefont {P.~C.}\ \bibnamefont {Main}}, \bibinfo {author}
  {\bibfnamefont {M.}~\bibnamefont {Henini}}, \bibinfo {author} {\bibfnamefont
  {C.~H.}\ \bibnamefont {Marrows}}, \ and\ \bibinfo {author} {\bibfnamefont
  {M.~A.}\ \bibnamefont {Howson}},\ }\href@noop {} {\bibfield  {journal}
  {\bibinfo  {journal} {Appl.Phys.Lett.}\ }\textbf {\bibinfo {volume} {74}},\
  \bibinfo {pages} {2507} (\bibinfo {year} {1999})}\BibitemShut {NoStop}%
\bibitem [{\citenamefont {Van\v{c}ura}\ \emph {et~al.}(2000)\citenamefont
  {Van\v{c}ura}, \citenamefont {Ihn}, \citenamefont {Broderick}, \citenamefont
  {Ensslin}, \citenamefont {Wegscheider},\ and\ \citenamefont
  {Bichler}}]{Vancura2000}%
  \BibitemOpen
  \bibfield  {author} {\bibinfo {author} {\bibfnamefont {T.}~\bibnamefont
  {Van\v{c}ura}}, \bibinfo {author} {\bibfnamefont {T.}~\bibnamefont {Ihn}},
  \bibinfo {author} {\bibfnamefont {S.}~\bibnamefont {Broderick}}, \bibinfo
  {author} {\bibfnamefont {K.}~\bibnamefont {Ensslin}}, \bibinfo {author}
  {\bibfnamefont {W.}~\bibnamefont {Wegscheider}}, \ and\ \bibinfo {author}
  {\bibfnamefont {M.}~\bibnamefont {Bichler}},\ }\href@noop {} {\bibfield
  {journal} {\bibinfo  {journal} {Phys.Rev.B}\ }\textbf {\bibinfo {volume}
  {62}},\ \bibinfo {pages} {5074} (\bibinfo {year} {2000})}\BibitemShut
  {NoStop}%
\bibitem [{\citenamefont {Ramezani~Masir}\ \emph {et~al.}(2008)\citenamefont
  {Ramezani~Masir}, \citenamefont {Vasilopoulos}, \citenamefont {Matulis},\
  and\ \citenamefont {Peeters}}]{Ramezani2008}%
  \BibitemOpen
  \bibfield  {author} {\bibinfo {author} {\bibfnamefont {M.}~\bibnamefont
  {Ramezani~Masir}}, \bibinfo {author} {\bibfnamefont {P.}~\bibnamefont
  {Vasilopoulos}}, \bibinfo {author} {\bibfnamefont {A.}~\bibnamefont
  {Matulis}}, \ and\ \bibinfo {author} {\bibfnamefont {F.~M.}\ \bibnamefont
  {Peeters}},\ }\href@noop {} {\bibfield  {journal} {\bibinfo  {journal}
  {Phys.Rev.B}\ }\textbf {\bibinfo {volume} {77}},\ \bibinfo {pages} {235443}
  (\bibinfo {year} {2008})}\BibitemShut {NoStop}%
\bibitem [{\citenamefont {Sharma}\ and\ \citenamefont
  {Ghosh}(2011)}]{Sharma2011}%
  \BibitemOpen
  \bibfield  {author} {\bibinfo {author} {\bibfnamefont {M.}~\bibnamefont
  {Sharma}}\ and\ \bibinfo {author} {\bibfnamefont {S.}~\bibnamefont {Ghosh}},\
  }\href@noop {} {\bibfield  {journal} {\bibinfo  {journal} {J.Phys.Cond.Mat.}\
  }\textbf {\bibinfo {volume} {23}},\ \bibinfo {pages} {055501} (\bibinfo
  {year} {2011})}\BibitemShut {NoStop}%
\bibitem [{\citenamefont {Haldane}\ and\ \citenamefont
  {Arovas}(1995)}]{Haldane1995}%
  \BibitemOpen
  \bibfield  {author} {\bibinfo {author} {\bibfnamefont {F.~D.~M.}\
  \bibnamefont {Haldane}}\ and\ \bibinfo {author} {\bibfnamefont {D.~P.}\
  \bibnamefont {Arovas}},\ }\href@noop {} {\bibfield  {journal} {\bibinfo
  {journal} {Phys.Rev.B}\ }\textbf {\bibinfo {volume} {52}},\ \bibinfo {pages}
  {4223} (\bibinfo {year} {1995})}\BibitemShut {NoStop}%
\bibitem [{\citenamefont {Young}\ \emph {et~al.}(2014)\citenamefont {Young},
  \citenamefont {Sanchez-Yamagishi}, \citenamefont {Hunt}, \citenamefont
  {Choi}, \citenamefont {Watanabe}, \citenamefont {Taniguchi}, \citenamefont
  {Ashoori},\ and\ \citenamefont {Jarillo-Herrero}}]{Young2014}%
  \BibitemOpen
  \bibfield  {author} {\bibinfo {author} {\bibfnamefont {A.~F.}\ \bibnamefont
  {Young}}, \bibinfo {author} {\bibfnamefont {J.~D.}\ \bibnamefont
  {Sanchez-Yamagishi}}, \bibinfo {author} {\bibfnamefont {B.}~\bibnamefont
  {Hunt}}, \bibinfo {author} {\bibfnamefont {S.~H.}\ \bibnamefont {Choi}},
  \bibinfo {author} {\bibfnamefont {K.}~\bibnamefont {Watanabe}}, \bibinfo
  {author} {\bibfnamefont {T.}~\bibnamefont {Taniguchi}}, \bibinfo {author}
  {\bibfnamefont {R.~C.}\ \bibnamefont {Ashoori}}, \ and\ \bibinfo {author}
  {\bibfnamefont {P.}~\bibnamefont {Jarillo-Herrero}},\ }\href@noop {}
  {\bibfield  {journal} {\bibinfo  {journal} {Nature}\ }\textbf {\bibinfo
  {volume} {505}},\ \bibinfo {pages} {528} (\bibinfo {year}
  {2014})}\BibitemShut {NoStop}%
\bibitem [{\citenamefont {Tarasov}\ \emph {et~al.}(2010)\citenamefont
  {Tarasov}, \citenamefont {Hugger}, \citenamefont {Xu}, \citenamefont
  {Cerchez}, \citenamefont {Heinzel}, \citenamefont {Zozoulenko}, \citenamefont
  {Gasser-Szerer}, \citenamefont {Reuter},\ and\ \citenamefont
  {Wieck}}]{Tarasov2010}%
  \BibitemOpen
  \bibfield  {author} {\bibinfo {author} {\bibfnamefont {A.}~\bibnamefont
  {Tarasov}}, \bibinfo {author} {\bibfnamefont {S.}~\bibnamefont {Hugger}},
  \bibinfo {author} {\bibfnamefont {H.~Y.}\ \bibnamefont {Xu}}, \bibinfo
  {author} {\bibfnamefont {M.}~\bibnamefont {Cerchez}}, \bibinfo {author}
  {\bibfnamefont {T.}~\bibnamefont {Heinzel}}, \bibinfo {author} {\bibfnamefont
  {I.~V.}\ \bibnamefont {Zozoulenko}}, \bibinfo {author} {\bibfnamefont
  {U.}~\bibnamefont {Gasser-Szerer}}, \bibinfo {author} {\bibfnamefont
  {D.}~\bibnamefont {Reuter}}, \ and\ \bibinfo {author} {\bibfnamefont {A.~D.}\
  \bibnamefont {Wieck}},\ }\href@noop {} {\bibfield  {journal} {\bibinfo
  {journal} {Phys.Rev.Lett.}\ }\textbf {\bibinfo {volume} {104}},\ \bibinfo
  {pages} {186801} (\bibinfo {year} {2010})}\BibitemShut {NoStop}%
\bibitem [{\citenamefont {Schuler}\ \emph {et~al.}(2014)\citenamefont
  {Schuler}, \citenamefont {Cerchez}, \citenamefont {Xu}, \citenamefont
  {Schluck},\ and\ \citenamefont {Heinzel}}]{Schuler2014}%
  \BibitemOpen
  \bibfield  {author} {\bibinfo {author} {\bibfnamefont {B.}~\bibnamefont
  {Schuler}}, \bibinfo {author} {\bibfnamefont {M.}~\bibnamefont {Cerchez}},
  \bibinfo {author} {\bibfnamefont {H.}~\bibnamefont {Xu}}, \bibinfo {author}
  {\bibfnamefont {J.}~\bibnamefont {Schluck}}, \ and\ \bibinfo {author}
  {\bibfnamefont {T.}~\bibnamefont {Heinzel}},\ }\href@noop {} {\bibfield
  {journal} {\bibinfo  {journal} {Phys.Rev.B}\ }\textbf {\bibinfo {volume}
  {90}},\ \bibinfo {pages} {201111} (\bibinfo {year} {2014})}\BibitemShut
  {NoStop}%
\bibitem [{\citenamefont {Hasan}\ and\ \citenamefont {Kane}(2010)}]{Hasan2010}%
  \BibitemOpen
  \bibfield  {author} {\bibinfo {author} {\bibfnamefont {M.~Z.}\ \bibnamefont
  {Hasan}}\ and\ \bibinfo {author} {\bibfnamefont {C.~L.}\ \bibnamefont
  {Kane}},\ }\href@noop {} {\bibfield  {journal} {\bibinfo  {journal} {Rev.
  Mod. Phys.}\ }\textbf {\bibinfo {volume} {82}},\ \bibinfo {pages} {3045}
  (\bibinfo {year} {2010})}\BibitemShut {NoStop}%
\bibitem [{\citenamefont {Xu}\ \emph {et~al.}(2015)\citenamefont {Xu},
  \citenamefont {Belopolski}, \citenamefont {Alidoust}, \citenamefont
  {Neupane}, \citenamefont {Bian}, \citenamefont {Zhang}, \citenamefont
  {Sankar}, \citenamefont {Chang}, \citenamefont {Yuan}, \citenamefont {Lee},
  \citenamefont {Huang}, \citenamefont {Zheng}, \citenamefont {Ma},
  \citenamefont {Sanchez}, \citenamefont {Wang}, \citenamefont {Bansil},
  \citenamefont {Chou}, \citenamefont {Shibayev}, \citenamefont {Lin},
  \citenamefont {Jia},\ and\ \citenamefont {Zahid~Hasan}}]{Xu2016}%
  \BibitemOpen
  \bibfield  {author} {\bibinfo {author} {\bibfnamefont {S.-Y.}\ \bibnamefont
  {Xu}}, \bibinfo {author} {\bibfnamefont {I.}~\bibnamefont {Belopolski}},
  \bibinfo {author} {\bibfnamefont {N.}~\bibnamefont {Alidoust}}, \bibinfo
  {author} {\bibfnamefont {M.}~\bibnamefont {Neupane}}, \bibinfo {author}
  {\bibfnamefont {G.}~\bibnamefont {Bian}}, \bibinfo {author} {\bibfnamefont
  {C.}~\bibnamefont {Zhang}}, \bibinfo {author} {\bibfnamefont
  {R.}~\bibnamefont {Sankar}}, \bibinfo {author} {\bibfnamefont
  {G.}~\bibnamefont {Chang}}, \bibinfo {author} {\bibfnamefont
  {Z.}~\bibnamefont {Yuan}}, \bibinfo {author} {\bibfnamefont {C.-C.}\
  \bibnamefont {Lee}}, \bibinfo {author} {\bibfnamefont {S.-M.}\ \bibnamefont
  {Huang}}, \bibinfo {author} {\bibfnamefont {H.}~\bibnamefont {Zheng}},
  \bibinfo {author} {\bibfnamefont {J.}~\bibnamefont {Ma}}, \bibinfo {author}
  {\bibfnamefont {D.~S.}\ \bibnamefont {Sanchez}}, \bibinfo {author}
  {\bibfnamefont {B.}~\bibnamefont {Wang}}, \bibinfo {author} {\bibfnamefont
  {A.}~\bibnamefont {Bansil}}, \bibinfo {author} {\bibfnamefont
  {F.}~\bibnamefont {Chou}}, \bibinfo {author} {\bibfnamefont {P.~P.}\
  \bibnamefont {Shibayev}}, \bibinfo {author} {\bibfnamefont {H.}~\bibnamefont
  {Lin}}, \bibinfo {author} {\bibfnamefont {S.}~\bibnamefont {Jia}}, \ and\
  \bibinfo {author} {\bibfnamefont {M.}~\bibnamefont {Zahid~Hasan}},\
  }\href@noop {} {\bibfield  {journal} {\bibinfo  {journal} {Science}\ }\textbf
  {\bibinfo {volume} {349}},\ \bibinfo {pages} {613} (\bibinfo {year}
  {2015})}\BibitemShut {NoStop}%
\bibitem [{\citenamefont {Shi}\ \emph {et~al.}(2002)\citenamefont {Shi},
  \citenamefont {Peeters}, \citenamefont {Edmonds},\ and\ \citenamefont
  {Gallagher}}]{Shi2002}%
  \BibitemOpen
  \bibfield  {author} {\bibinfo {author} {\bibfnamefont {J.}~\bibnamefont
  {Shi}}, \bibinfo {author} {\bibfnamefont {F.~M.}\ \bibnamefont {Peeters}},
  \bibinfo {author} {\bibfnamefont {K.~W.}\ \bibnamefont {Edmonds}}, \ and\
  \bibinfo {author} {\bibfnamefont {B.~L.}\ \bibnamefont {Gallagher}},\
  }\href@noop {} {\bibfield  {journal} {\bibinfo  {journal} {Phys.Rev.B}\
  }\textbf {\bibinfo {volume} {66}},\ \bibinfo {pages} {035328} (\bibinfo
  {year} {2002})}\BibitemShut {NoStop}%
\bibitem [{\citenamefont {Charbonneau}\ \emph {et~al.}(1982)\citenamefont
  {Charbonneau}, \citenamefont {van Vliet},\ and\ \citenamefont
  {Vasilopoulos}}]{Charbonneau1982}%
  \BibitemOpen
  \bibfield  {author} {\bibinfo {author} {\bibfnamefont {M.}~\bibnamefont
  {Charbonneau}}, \bibinfo {author} {\bibfnamefont {K.~M.}\ \bibnamefont {van
  Vliet}}, \ and\ \bibinfo {author} {\bibfnamefont {P.}~\bibnamefont
  {Vasilopoulos}},\ }\href@noop {} {\bibfield  {journal} {\bibinfo  {journal}
  {J. Math. Phys.}\ }\textbf {\bibinfo {volume} {23}},\ \bibinfo {pages} {318}
  (\bibinfo {year} {1982})}\BibitemShut {NoStop}%
\bibitem [{\citenamefont {Thornton}\ \emph {et~al.}(1989)\citenamefont
  {Thornton}, \citenamefont {Roukes}, \citenamefont {Scherer},\ and\
  \citenamefont {Van~de Gaag}}]{Thornton1989}%
  \BibitemOpen
  \bibfield  {author} {\bibinfo {author} {\bibfnamefont {T.~J.}\ \bibnamefont
  {Thornton}}, \bibinfo {author} {\bibfnamefont {M.~L.}\ \bibnamefont
  {Roukes}}, \bibinfo {author} {\bibfnamefont {A.}~\bibnamefont {Scherer}}, \
  and\ \bibinfo {author} {\bibfnamefont {B.~P.}\ \bibnamefont {Van~de Gaag}},\
  }\href@noop {} {\bibfield  {journal} {\bibinfo  {journal} {Phys.Rev.Lett.}\
  }\textbf {\bibinfo {volume} {63}},\ \bibinfo {pages} {2128} (\bibinfo {year}
  {1989})}\BibitemShut {NoStop}%
\bibitem [{SM()}]{SM}%
  \BibitemOpen
  \href@noop {} {}\bibinfo {note} {See Supplemental Material at [URL inserted
  by publisher] for device micrographs, a detailed calculation of the channel
  width and of the conductivity tensor}\BibitemShut {NoStop}%
\bibitem [{\citenamefont {Fang}\ and\ \citenamefont {Howard}(1966)}]{Fang1966}%
  \BibitemOpen
  \bibfield  {author} {\bibinfo {author} {\bibfnamefont {F.~F.}\ \bibnamefont
  {Fang}}\ and\ \bibinfo {author} {\bibfnamefont {W.~E.}\ \bibnamefont
  {Howard}},\ }\href@noop {} {\bibfield  {journal} {\bibinfo  {journal}
  {Phys.Rev.Lett.}\ }\textbf {\bibinfo {volume} {16}},\ \bibinfo {pages} {797}
  (\bibinfo {year} {1966})}\BibitemShut {NoStop}%
\bibitem [{\citenamefont {Edmonds}\ \emph {et~al.}(2001)\citenamefont
  {Edmonds}, \citenamefont {Gallagher}, \citenamefont {Main}, \citenamefont
  {Overend}, \citenamefont {Wirtz}, \citenamefont {Nogaret}, \citenamefont
  {Henini}, \citenamefont {Marrows}, \citenamefont {Hickey},\ and\
  \citenamefont {Thoms}}]{Edmonds2001}%
  \BibitemOpen
  \bibfield  {author} {\bibinfo {author} {\bibfnamefont {K.~W.}\ \bibnamefont
  {Edmonds}}, \bibinfo {author} {\bibfnamefont {B.~L.}\ \bibnamefont
  {Gallagher}}, \bibinfo {author} {\bibfnamefont {P.~C.}\ \bibnamefont {Main}},
  \bibinfo {author} {\bibfnamefont {N.}~\bibnamefont {Overend}}, \bibinfo
  {author} {\bibfnamefont {R.}~\bibnamefont {Wirtz}}, \bibinfo {author}
  {\bibfnamefont {A.}~\bibnamefont {Nogaret}}, \bibinfo {author} {\bibfnamefont
  {M.}~\bibnamefont {Henini}}, \bibinfo {author} {\bibfnamefont {C.~H.}\
  \bibnamefont {Marrows}}, \bibinfo {author} {\bibfnamefont {B.~J.}\
  \bibnamefont {Hickey}}, \ and\ \bibinfo {author} {\bibfnamefont
  {S.}~\bibnamefont {Thoms}},\ }\href@noop {} {\bibfield  {journal} {\bibinfo
  {journal} {Phys.Rev.B}\ }\textbf {\bibinfo {volume} {64}},\ \bibinfo {pages}
  {041303} (\bibinfo {year} {2001})}\BibitemShut {NoStop}%
\bibitem [{\citenamefont {Reed}\ \emph {et~al.}(1988)\citenamefont {Reed},
  \citenamefont {Randall}, \citenamefont {Aggarwal}, \citenamefont {Matyi},
  \citenamefont {Moore},\ and\ \citenamefont {Wetsel}}]{Reed1988}%
  \BibitemOpen
  \bibfield  {author} {\bibinfo {author} {\bibfnamefont {M.~A.}\ \bibnamefont
  {Reed}}, \bibinfo {author} {\bibfnamefont {J.~N.}\ \bibnamefont {Randall}},
  \bibinfo {author} {\bibfnamefont {R.~J.}\ \bibnamefont {Aggarwal}}, \bibinfo
  {author} {\bibfnamefont {R.~J.}\ \bibnamefont {Matyi}}, \bibinfo {author}
  {\bibfnamefont {T.~M.}\ \bibnamefont {Moore}}, \ and\ \bibinfo {author}
  {\bibfnamefont {A.~E.}\ \bibnamefont {Wetsel}},\ }\href@noop {} {\bibfield
  {journal} {\bibinfo  {journal} {Phys.Rev.Lett.}\ }\textbf {\bibinfo {volume}
  {60}},\ \bibinfo {pages} {535} (\bibinfo {year} {1988})}\BibitemShut
  {NoStop}%
\bibitem [{\citenamefont {Press}\ \emph {et~al.}(1994)\citenamefont {Press},
  \citenamefont {Teukolsky}, \citenamefont {Vetterling},\ and\ \citenamefont
  {Flannery}}]{Press1994}%
  \BibitemOpen
  \bibfield  {author} {\bibinfo {author} {\bibfnamefont {W.~H.}\ \bibnamefont
  {Press}}, \bibinfo {author} {\bibfnamefont {S.~A.}\ \bibnamefont
  {Teukolsky}}, \bibinfo {author} {\bibfnamefont {W.~T.}\ \bibnamefont
  {Vetterling}}, \ and\ \bibinfo {author} {\bibfnamefont {B.~P.}\ \bibnamefont
  {Flannery}},\ }\href@noop {} {\emph {\bibinfo {title} {Numerical Recipes in
  C}}}\ (\bibinfo  {publisher} {CUP},\ \bibinfo {year} {1994})\BibitemShut
  {NoStop}%
\bibitem [{\citenamefont {Pippard}(1989)}]{Pippard1989}%
  \BibitemOpen
  \bibfield  {author} {\bibinfo {author} {\bibfnamefont {A.~B.}\ \bibnamefont
  {Pippard}},\ }\href@noop {} {\emph {\bibinfo {title} {Magnetoresistance in
  metals}}}\ (\bibinfo  {publisher} {CUP},\ \bibinfo {year} {1989})\BibitemShut
  {NoStop}%
\bibitem [{\citenamefont {Chklowskii}\ \emph {et~al.}(1992)\citenamefont
  {Chklowskii}, \citenamefont {Schklovskii},\ and\ \citenamefont
  {Glazman}}]{Chklovskii1992}%
  \BibitemOpen
  \bibfield  {author} {\bibinfo {author} {\bibfnamefont {D.~B.}\ \bibnamefont
  {Chklowskii}}, \bibinfo {author} {\bibfnamefont {B.~I.}\ \bibnamefont
  {Schklovskii}}, \ and\ \bibinfo {author} {\bibfnamefont {L.~I.}\ \bibnamefont
  {Glazman}},\ }\href@noop {} {\bibfield  {journal} {\bibinfo  {journal}
  {Phys.Rev.B}\ }\textbf {\bibinfo {volume} {46}},\ \bibinfo {pages} {4026}
  (\bibinfo {year} {1992})}\BibitemShut {NoStop}%
\bibitem [{\citenamefont {Siddiki}\ and\ \citenamefont
  {Gerhardts}(2004)}]{Siddiki2004}%
  \BibitemOpen
  \bibfield  {author} {\bibinfo {author} {\bibfnamefont {A.}~\bibnamefont
  {Siddiki}}\ and\ \bibinfo {author} {\bibfnamefont {R.~R.}\ \bibnamefont
  {Gerhardts}},\ }\href@noop {} {\bibfield  {journal} {\bibinfo  {journal}
  {Phys.Rev.B}\ }\textbf {\bibinfo {volume} {70}},\ \bibinfo {pages} {195335}
  (\bibinfo {year} {2004})}\BibitemShut {NoStop}%
\bibitem [{\citenamefont {Potter}\ \emph {et~al.}(2014)\citenamefont {Potter},
  \citenamefont {Kimchi},\ and\ \citenamefont {Vishwanath}}]{Potter2014}%
  \BibitemOpen
  \bibfield  {author} {\bibinfo {author} {\bibfnamefont {A.~C.}\ \bibnamefont
  {Potter}}, \bibinfo {author} {\bibfnamefont {I.}~\bibnamefont {Kimchi}}, \
  and\ \bibinfo {author} {\bibfnamefont {A.}~\bibnamefont {Vishwanath}},\
  }\href@noop {} {\bibfield  {journal} {\bibinfo  {journal} {Nature
  Communications}\ }\textbf {\bibinfo {volume} {5}},\ \bibinfo {pages} {5161}
  (\bibinfo {year} {2014})}\BibitemShut {NoStop}%
\end{thebibliography}%
\bibliographystyle{apsrev4-1}

\end{document}